\documentclass[notitlepage,english,aps,floats,onecolumn,showpacs,nofootinbib,floatfix]{revtex4-1}

\usepackage{pslatex}
\usepackage[T1]{fontenc}
\usepackage[latin1]{inputenc}
\usepackage{graphicx}
\usepackage{epsfig}
\usepackage{longtable}
\usepackage{float}
\usepackage{calc}
\usepackage{ifthen}
\usepackage{amsmath}
\usepackage{hyperref}
\usepackage{amssymb}

\usepackage{color}

{
{
{
\newcommand{\bea}{\begin{eqnarray}}
\newcommand{\eea}{\end{eqnarray}}

\newcommand{\nc}{\newcommand}
\nc{\renc}{\renewcommand}
\nc{\eqs}[2]{\mbox{Eqs.~(\ref{#1},\,\ref{#2})}}
\nc{\eq}[1]{\mbox{Eq.~(\ref{#1})}}
\nc{\figs}[2]{\mbox{Figs.~(\ref{#1},\,\ref{#2})}}
\nc{\fig}[1]{\mbox{Fig~.(\ref{#1})}}
\nc{\be}[1]{\begin{equation} \mbox{$\label{#1}$}}
\nc{\ee}{\vspace{0.1cm}\end{equation}}

\newcommand{\bean}{\begin{eqnarray*}}
\newcommand{\eean}{\end{eqnarray*}}

\def\GeV{{\rm \ GeV}}

\def\lae{\;^{<}_{\sim} \;} \def\gae{\; ^{>}_{\sim} \;}

\begin{document}

\title{Higgs Inflation Using the Unstable Standard Model  Potential}

\author{J.McDonald }
\email{j.mcdonald@lancaster.ac.uk}
\affiliation{Dept. of Physics,  
Lancaster University, Lancaster LA1 4YB, UK}

\begin{abstract}

It is likely that the Higgs potential of the Standard Model is unstable, turning negative at $\phi < \Lambda \sim 10^{10}$ GeV. Here we consider whether it is possible to have Higgs Inflation on the positive stable region of the potential at $\phi < \Lambda$. To do this we add a non-minimally coupled induced gravity sector with scalar $\chi$ to the Standard Model. For an appropriate form for the non-minimal coupling of $\chi$, we show that it is possible to have conventional Higgs inflation at small $\phi < \Lambda$ if the effective Planck mass in the Jordan frame during inflation is sufficiently small, with a phase transition to $\chi \neq 0$ at the end of Higgs inflation which increases the Jordan frame Planck mass to its presently observed value. In the Einstein frame this corresponds to a suppression of the Higgs kinetic and potential term at the end of inflation. We show that the predictions of Higgs inflation at tree level are unaltered from conventional Higgs Inflation, with the exception of the magnitude of the Higgs field during inflation. Hence Higgs Inflation can be achieved using the potential of the unmodified Standard Model.

\end{abstract}
 \pacs{}
 
\maketitle

\section{Introduction}   

It is believed that the electroweak vacuum of the Standard Model (SM) is metastable, due to quantum corrections which cause the Higgs effective potential to become negative at an instability scale $\phi = \Lambda \sim 10^{10} \GeV$ \cite{unst1,unst2,unst3}. (The metastability of the SM Higgs potential has a long history; see
 \cite{frog} for an early discussion of the issue and \cite{sher} for a review.) Such a potential, if unmodified at large $\phi$, could not serve as a basis for non-minimally coupled Higgs Inflation \cite{hi}.

It is possible that the potential could be modified to overcome this; for example by adding particles to the Standard Model to modify the quantum corrections, such as TeV scalars with a sufficiently strong portal coupling to the Higgs 
\cite{st1,st2,st3,fixcorr1,fixcorr2}. (For a recent review of stabilisation mechanisms, see \cite{lit}.) However, it may be the SM Higgs potential is indeed unstable, as would be the case if such new particles either do not exist or are too weakly coupled to the Standard Model to sufficiently modify the potential. Here we propose that it may be possible to use the "healthy" positive part of the unmodified Standard Model Higgs potential at $\phi \lae \Lambda$ to support Higgs Inflation, if the effective Planck mass in the Jordan frame, $M_{Pl\;eff}$, is much smaller during inflation than it is at present. As a specific example, we will show that this can be achieved in a model with an induced gravity\footnote{"Induced gravity" in the sense of \cite{ig1,ig2}, where the Planck mass is determined by the expectation value of a scalar field.} scalar $\chi$ that is non-minimally coupled to gravity and whose vacuum expectation value determines the present value of the Planck mass. With an appropriate form for the non-minimal coupling of $\chi$, it is possible for the induced gravity phase transition from $\chi = 0$ to occur at the end of Higgs Inflation, increasing the effective Planck mass to its present value. In the Einstein frame, the Planck mass is fixed to its present value and the mechanism of the model becomes the suppression of the Higgs  kinetic and potential term by the increase of $\chi$ at the end of inflation. We will show that the resulting model leaves the classical  predictions of Higgs Inflation for the spectral index, $n_{s}$, the tensor-to-scalar ratio, $r$, and the reheating temperature, $T_{R}$, unchanged, but can reduce $\phi$ during inflation to values below the Higgs instability scale.    

\section{The Model} 

 We will consider conventional metric Higgs Inflation\footnote{Palatini Higgs Inflation is a viable alternative \cite{pal}.}. The Higgs potential $V(\phi)$ at $\phi \gg v$, where $v$ the present Higgs vacuum expectation value, is a quartic potential with a running Higgs self-coupling $\lambda_{\phi}(\mu)$, where $\mu$ may be chosen to equal\footnote{The renormalisation of non-minimally coupled models depends upon the conformal frame in which the renormalisation is performed. Whilst there are proposals for a preferred frame \cite{george}, it has also been proposed that the correct frame can only be determined from an ultra-violet completion of the model \cite{uvc}. For values of $\phi$ for which the conformal factor $\Omega \approx 1$, the Jordan and Einstein frames are indistinguishable from the Standard Model, therefore the quantum corrections are conventional Standard Model corrections and $\mu = \phi$ is a good choice for the renormalisation scale. For larger field values, the quantum corrections to the potential and the best choice for $\mu$  become frame dependent \cite{pt1c}.} $\phi$. In this discussion we will consider $\phi$ to be small compared to the instability scale and so $\lambda_{\phi} > 0$. The action of the model in the Jordan frame is 
\be{e1}      S = \int d^{4} x \sqrt{-g} \left[ \left( M_{0}^{2} + \xi_{\phi} \phi^{2} + G(\chi) \right) \frac{R}{2}   - \frac{1}{2} \partial_{\mu} \phi \partial^{\mu} \phi - \frac{1}{2} \partial_{\mu} \chi \partial^{\mu} \chi - V(\phi, \chi) \right]  ~,\ee
where $M_{0} \ll M_{Pl}$ and  
\be{e2}  V(\phi, \chi) =   \frac{\lambda_{\phi} \phi^4}{4} + \frac{\lambda_{\chi}}{4} \left( \chi^{2} - v_{\chi}^{2} \right)^{2}    ~.\ee
In the following we define the effective Planck mass in the Jordan frame, $M_{Pl\;eff}$, to be 
\be{e3}  M_{Pl\;eff} =  \left( M_{0}^{2} + \xi_{\phi} \phi^{2} + G(\chi) \right)^{1/2}  ~.\ee
Here $G(\chi)$ defines the non-minimal coupling of $\chi$. For the model to work, $G(\chi)$ should be negative at small $\chi$ and positive at large $\chi$. Assuming that $G(\chi)$ is a function of $\chi^2$, the leading order term in the expansion of $G(\chi)$ at small $\chi$ should therefore be of the form 
\be{e4}   G(\chi) \approx   - \xi_{\chi} \chi^{2}  \;\;;\;\; \xi_{\chi} > 0 ~,\ee  

In order to analyse inflation, we transform to the Einstein frame\footnote{After inflation $\Omega = 1$, therefore results for observables calculated in the Jordan and Einstein frames are indistinguishable.}.  We first write \eq{e1} in the form 
\be{e4a}       S = \int d^{4} x \sqrt{-g} \left[  \frac{M_{Pl}^{2}}{2} \Omega^{2} R   - \frac{1}{2} \partial_{\mu} \phi \partial^{\mu} \phi - \frac{1}{2} \partial_{\mu} \chi \partial^{\mu} \chi - V(\phi, \chi) \right]  ~,\ee
where the conformal factor $\Omega$ is defined by
\be{e5} \Omega^{2} = \frac{M_{0}^{2}}{M_{Pl}^{2}} + \frac{\xi_{\phi} \phi^{2}}{M_{Pl}^{2}} + \frac{G(\chi)}{M_{Pl}^{2}}   ~.\ee
The Einstein frame action is then 
\be{e6} S = \int d^{4} x \sqrt{-\tilde{g}} \left[ \frac{M_{Pl}^{2}}{2} \tilde{R} - \frac{3 M_{Pl}^{2}}{4 \Omega^{4}}{\partial_{\mu} \Omega^{2} \partial^{\mu} \Omega^{2} } - \frac{1}{2 \Omega^{2}} \partial_{\mu}\phi  \partial^{\mu}\phi   - \frac{1}{2 \Omega^{2}} \partial_{\mu}\chi  \partial^{\mu}\chi - \frac{V(\phi, \chi)}{\Omega^{4}}   \right]   ~,\ee
where the metric is $\tilde{g}_{\mu \nu} =  \Omega^{2} g_{\mu \nu}$.   
After expanding the second term in \eq{e6}, the kinetic terms are  
\be{e7} -\frac{1}{2 \Omega^{2}} \left(1 + \frac{6 \xi_{\phi}^{2} \phi^{2} }{\Omega^{2} M_{Pl}^{2}} \right) \partial_{\mu} \phi \partial^{\mu} \phi 
- \frac{3 \xi_{\phi}}{\Omega^{4} M_{Pl}^{2}} G'(\chi) \phi \partial_{\mu} \chi \partial^{\mu} \phi 
 -\frac{1}{2 \Omega^{2}} \left(1 + \frac{3 G'(\chi)^{2} }{2 \Omega^{2} M_{Pl}^{2}} \right) \partial_{\mu} \chi \partial^{\mu} \chi    ~,\ee
where $G'(\chi) = d G/d \chi$. 

We consider inflation in the $\phi$ direction. We assume that $\chi = 0$ during inflation and later show that this can be ensured by the non-minimal coupling \eq{e4}. In the limit of small $\chi$, the conformal factor becomes 
\be{e8} \Omega^{2} = \frac{M_{0}^{2}}{M_{Pl}^{2}} + \frac{\xi_{\phi} \phi^{2}}{M_{Pl}^{2}} - \frac{\xi_{\chi} \chi^{2}}{M_{Pl}^{2}}  ~.\ee
It will be convenient to define $\Omega^{2}$ in the form
$\Omega^{2} = (M_{0}/M_{Pl})^{2} \tilde{\Omega}^{2}$, 
where 
\be{e9} \tilde{\Omega}^{2} = 1 + \frac{\xi_{\phi} \phi^{2}}{M_{0}^{2}} - \frac{\xi_{\chi} \chi^{2}}{M_{0}^{2}}  ~.\ee
The $\phi$ kinetic term with $\chi = 0$ can then be written as  
\be{e10} -\frac{1}{2 \tilde{\Omega}^{2}} \left( 1 + \frac{6 \xi_{\phi}^{2} \tilde{\phi}^{2}}{\tilde{\Omega}^{2} M_{Pl}^{2} } \right) \partial_{\mu} \tilde{\phi}  \partial^{\mu} \tilde{\phi} \;\;\;;\;\;\; \tilde{\Omega}^{2} = 1 + \frac{\xi_{\phi} \tilde{\phi}^{2} }{ M_{Pl}^{2}}  ~,\ee
where we define $\tilde{\phi} = (M_{Pl}/M_{0}) \phi$.  
The Einstein frame Higgs potential in terms of $\tilde{\phi}$ and $\tilde{\Omega}$ is then 
\be{e12}  V_{E}(\phi) = \frac{V(\phi)}{\Omega^{4}} = \frac{\lambda_{\phi} \phi^{4}}{4 \Omega^{4}}  
= \frac{\lambda_{\phi} \tilde{\phi}^{4}}{4 \tilde{\Omega}^{4}}
 ~.\ee  
Therefore, in terms of $\tilde{\phi}$, the action is exactly the same as that of conventional Higgs Inflation and therefore will produce the same predictions as a function of the number of e-folds $N$.  As a result, if the number of e-foldings corresponding to the pivot scale at horizon exit, $N_{*}$, has the same value as in conventional Higgs Inflation, then the predictions of the model for $n_{s}$ and $r$ will be the same as conventional Higgs Inflation. In addition, as in conventional Higgs Inflation, the Higgs field $\tilde{\phi}$ at $N$ e-foldings is \cite{hi}
\be{e15} \tilde{\phi} = \sqrt{\frac{4 N}{3 \xi_{\phi}}} M_{Pl}   ~\ee
and 
the curvature power spectrum is 
\be{e17} P_{R} = \frac{\lambda_{\phi} N^{2}}{72 \pi^{2} 
\xi_{\phi}^{2}}   ~. \ee 
The observed curvature power at the pivot scale, 
$P_{R} = 2.1 \times 10^{-9}$, then gives 
\be{e18} \xi_{\phi} = 8.2 \times 10^{2} \sqrt{\lambda_{\phi}} N_{*} ~.\ee 

At the end of inflation, $N \approx 1$ and $\tilde{\phi}_{e} = 2 M_{Pl}/\sqrt{3 \xi_{\phi}}$. Since this is only slightly different from the value $\tilde{\phi} = M_{Pl}/\sqrt{\xi_{\phi}}$ below which  $\Omega \approx M_{0}/M_{Pl}$ and $\tilde{\Omega} \approx 1$ when $\chi = 0$, for convenience in the following we will use $\tilde{\phi}_{e}  \approx M_{Pl}/\sqrt{\xi_{\phi}}$, $\tilde{\Omega}(\tilde{\phi}_{e}) \approx 1$ and  $\Omega(\tilde{\phi}_{e}) \approx M_{0}/M_{Pl}$ at the end of inflation. At this time $\chi = 0$ and the effective Planck mass in the Jordan frame is $M_{Pl \; eff} \approx M_{0}$. The energy density at the end of inflation in the Einstein frame is 
\be{e20}  \rho_{E} \approx \frac{\lambda_{\phi} M_{Pl}^{4}}{4 \xi_{\phi}^{2}}    ~,\ee
as in conventional Higgs Inflation.  
At the end of inflation, the $\phi$ field will enter into oscillations and decay to SM radiation and $\phi$ will relax to zero. 
As shown below, the $\chi = 0$ minimum will then become unstable and $\chi$ will evolve to its minimum at $\chi = v_{\chi}$. The conformal factor with $\phi = 0$ and $\chi = v_{\chi}$ becomes $\Omega = 1$ and the Jordan frame effective Planck mass becomes $M_{Pl}$. Assuming that the contribution to the potential energy at the end of inflation is dominated by $V(\phi)$ (we discuss this condition below), and assuming instant reheating via rapid Higgs field preheating \cite{pre}, the reheating temperature will be the same as in Higgs Inflation, $T_{R} \approx \rho_{E}^{1/4} \approx M_{Pl}/\sqrt{\xi_{\phi}}$. Moreover, since both the reheating temperature and the energy density, and hence horizon during inflation, are unchanged from Higgs Inflation, it follows that $N_{*}$ is also unchanged. Therefore the predictions of the model will be the same as in Higgs Inflation, except for the value of $\phi(N)$, which is suppressed by a factor $M_{0}/M_{Pl}$ relative to its value in conventional Higgs Inflation,  
\be{e21} \phi(N) = \left(\frac{M_{0}}{M_{Pl}}\right) \tilde{\phi}(N) = \sqrt{\frac{4 N}{3 \xi_{\phi}}} M_{0}   ~.\ee 
At the pivot scale, and using $\lambda_{\phi} \approx 0.1$ for the SM Higgs quartic coupling and \eq{e18} for $\xi_{\phi}$, the value of $\phi(N_{*})$ is numerically 
\be{e22} \phi(N_{*}) = 7.2 \times 10^{9} \left(\frac{0.1}{\lambda_{\phi}}\right)^{1/4} \left(\frac{M_{0}}{10^{11} \GeV}\right) \GeV    ~.\ee
Therefore $M_{0} \lae 10^{11} \GeV$ will allow Higgs  Inflation to take place on the positive part of the metastable Higgs potential at $\phi < \Lambda \approx 10^{10} \GeV$. 

\section{The Induced Gravity phase transition} 

We next consider how the scenario discussed in the previous section can be realised in practice, by $\chi$ becoming unstable and developing an expectation value once $\phi \rightarrow 0$ at the end of inflation. 
We first discuss how the $\chi = 0$ minimum can be stable during inflation due to the non-minimal coupling of $\chi$. We can write the $\chi$ potential in the form
\be{e40}  V(\chi) = V_{0} - \frac{1}{2} \mu_{\chi}^{2} \chi^{2} + \frac{\lambda_{\chi}}{4} \chi^{4}   ~,\ee
where $V_{0} = \mu_{\chi}^{4}/4 \lambda_{\chi}$ and
$\mu_{\chi}^{2} = \lambda_{\chi} v_{\chi}^{2} $. The full Einstein frame potential is then 
\be{e41}  V_{E}(\phi, \chi) =  \frac{ V(\phi) + V(\chi)}{\Omega^{4}} ~.\ee
For small $\chi$, this can be written as 
\be{e42}  V_{E}(\phi, \chi) \approx \frac{V_{0} + \frac{\lambda_{\phi}}{4} \phi^{4} - \frac{\mu_{\chi}^{2}}{2}  \chi^{2}  }{\left( \frac{M_{0}^{2}}{M_{Pl}^{2}} + \frac{\xi_{\phi} \phi^{2}}{M_{Pl}^{2}} - \frac{\xi_{\chi} \chi^{2}}{M_{Pl}^{2}} \right)^{2}  }   ~.\ee 
 We assume that $V_{0} \ll \lambda_{\phi} \phi^{4}/4$ during inflation, in which case we can neglect $V_{0}$ (we justify this below).  
During inflation, $\xi_{\phi} \phi^{2}/M_{Pl}^{2} > M_{0}^{2}/M_{Pl}^{2}$ and we can therefore approximate the potential as 
\be{e43} V_{E} \approx \frac{M_{Pl}^{4}}{\xi_{\phi}^{2} \phi^{4}} \frac{ \left(    
\frac{\lambda_{\phi}}{4} \phi^{4} - \frac{1}{2} \mu_{\chi}^{2} \chi^{2}\right) }{ 
\left(1 - \frac{\xi_{\chi} \chi^{2}}{\xi_{\phi} \phi^{2}} \right)^{2} }   ~\ee
For small $\chi$, we expand this as 
\be{e44} V_{E} \approx \frac{M_{Pl}^{4}}{\xi_{\phi}^{2} \phi^{4}}\left(\frac{\lambda_{\phi}}{4} \phi^{4} - \frac{1}{2} \mu_{\chi}^{2} \chi^{2}\right)\left(1 + 
 \frac{2 \xi_{\chi} \chi^{2}}{ \xi_{\phi} \phi^{2} } \right)  ~.\ee
Therefore the $\chi$ mass term in the inflaton background is 
\be{e45}  \frac{M_{Pl}^{4}}{\xi_{\phi}^{2} \phi^{4}} \left( \frac{\lambda_{\phi}}{2} \frac{\xi_{\chi}}{\xi_{\phi}} \phi^{2} - \frac{\mu_{\chi}^{2}}{2} \right) \chi^{2}   ~,\ee 
and the non-minimal coupling $\xi_{\chi}$ will stabilise the $\chi = 0$ minimum if 
\be{e46}   \xi_{\chi} > \frac{\mu_{\chi}^{2}}{\lambda_{\phi}} \frac{\xi_{\phi}}{\phi^{2}} ~\ee
Since $\phi > \phi_{e} = M_{0}/\sqrt{\xi_{\phi}}$ during inflation, this will be true throughout inflation if 
\be{e47} \mu_{\chi} \lae \left(\lambda_{\phi} \xi_{\chi}\right)^{1/2} \frac{M_{0}}{\xi_{\phi}}  = 2.2 \times 10^{8} \left(\frac{\xi_{\chi}}{10^{4}}\right)^{1/2}  \left(\frac{M_{0}}{10^{11} \GeV}\right) \GeV ~,\ee
where we set $N_{*} = 55$ in all of our numerical expressions, in which case $\xi_{\phi} = 4.5 \times 10^{4} \lambda_{\phi}^{1/2}$.

This gives the condition for classical stability of the $\chi = 0$ minimum during inflation. In addition, we need to check that the minimum is stable with respect to quantum fluctuations of $\chi$ during inflation. This will be true if the mass of the canonically normalised $\chi$ field is large compared to $H$ during inflation. During inflation, the kinetic term of the $\chi$ field in the inflaton background is 
\be{e48} -\frac{1}{2 \Omega^{2}} \partial_{\mu}\chi \partial^{\mu}\chi \equiv -\frac{1}{2} \partial_{\mu}\tilde{\chi} \partial^{\mu}\tilde{\chi}  ~,\ee
where $\Omega^{2} \approx \xi_{\phi} \phi^{2}/M_{Pl}^{2}$ and we define the canonically normalised scalar as $\tilde{\chi} = \chi/\Omega$. In terms of $\tilde{\chi}$, the mass term during inflation becomes 
\be{e49} \frac{M_{Pl}^{4}}{\xi_{\phi}^{2} \phi^{4}} \left( \frac{\lambda_{\phi} \xi_{\chi} \phi^{2}}{2 \xi_{\phi}} - \frac{\mu_{\chi}^{2}}{2}\right) \Omega^{2} \tilde{\chi}^{2}  = \frac{M_{Pl}^{2}}{\xi_{\phi} \phi^{2}} \left(  \frac{\lambda_{\phi} \xi_{\chi} \phi^{2}}{2 \xi_{\phi}}  - \frac{\mu_{\chi}^{2}}{2}\right) \tilde{\chi}^{2}
 ~.\ee 
Therefore, assuming that the positive contribution to the mass squared term dominates, the mass of the scalar $\tilde{\chi}$ is given by 
\be{e50} m_{\tilde{\chi}}^{2} \approx \frac{\lambda_{\phi} \xi_{\chi} M_{Pl}^{2}}{\xi_{\phi}^{2}} ~.\ee
During inflation in the Einstein frame, $H^{2} = \lambda_{\phi} M_{Pl}^{2}/12 \xi_{\phi}^{2} $. Therefore the condition for suppression of $\tilde{\chi}$ quantum fluctuations during inflation, $m_{\tilde{\chi}}^{2} \gg H^{2}$, is satisfied if 
\be{e51} \xi_{\chi} \gg \frac{1}{12}    ~.\ee
This is easily satisfied. Thus $\chi = 0$ during inflation can be both classically stable and safe with respect to $\tilde{\chi}$ quantum fluctuations during inflation. 

In the above we have assumed that the potential at the end of inflation is dominated by $V(\phi)$. This requires that 
\be{e52} \frac{\lambda_{\phi} \phi_{e}^{4}}{4} >  V_{0} = \frac{\mu_{\chi}^{4}}{4 \lambda_{\chi}} \Rightarrow \mu_{\chi} < \frac{ \left(\lambda_{\phi} \lambda_{\chi}\right)^{1/4} M_{0}}{\sqrt{\xi_{\phi}} }  = 2.7 \times 10^{8} \left(
\frac{\lambda_{\chi}}{0.1} \right)^{1/4} 
\left(\frac{M_{0}}{10^{11} \GeV} \right) \GeV ~.\ee
Since $M_{0} \lae 10^{11} \GeV$ in order to have $\phi < \Lambda\approx 10^{10} \GeV$ during inflation, it follows that $\mu_{\chi} \lae 10^{8} \GeV$ in this class of model. 

Finally, we are assuming that $\chi$ field can rapidly roll from the $\chi = 0$ minimum once $\phi = 0$. This requires that the magnitude of the negative mass squared of the canonically normalised field $\tilde{\chi}$ in the $\chi = \phi = 0$ minimum, $\tilde{\chi} = (M_{Pl}/M_{0}) \chi$, is greater than $H^{2}$. From \eq{e42}, the negative mass squared term in this case is  
\be{e53} -\left(\frac{M_{Pl}}{M_{0}}\right)^{2} \frac{\mu_{\chi}^{2}}{2} \tilde{\chi}^{2}   ~.\ee
The magnitude of the negative mass squared term is therefore large compared to $H^2$ at the end of inflation if
\be{e54} \mu_{\chi} \gae \sqrt{\frac{\lambda_{\phi}}{12} } \frac{M_{0}}{\xi_{\phi}} = 6.4 \times 10^{5} \left(\frac{M_{0}}{10^{11} \GeV} \right) \GeV  ~.\ee
We note that this constraint it not as  essential as the others, as the $\chi$ transition will in any case eventually occur once $H$ decreases sufficiently due to expansion. 

An example of model parameters for which the potential is dominated by $V(\phi)$ during inflation, $\phi$ during inflation is less than $\Lambda \sim 10^{10} \GeV$, $\chi = 0$  is stable during inflation, and the $\chi$ transition occurs once $\phi \rightarrow 0$ at the end of inflation, is given by $\mu_{\chi} = 10^{6} \GeV$, $M_{0} = 10^{10} \GeV$, $\xi_{\chi} = 10^{4}$, and $\lambda_{\chi} = \lambda_{\phi} = 0.1$, with $\phi(N = 55) = 7.2 \times 10^{8} \GeV$.

\section{The $\chi$ Non-minimal Coupling}  

The condition that the potential is dominated by the Higgs potential during inflation implies that $\mu_{\chi} \lae 10^{8} \GeV$ and so $v_{\chi} = \mu_{\chi}/\sqrt{\lambda_{\chi}}$ will be much smaller than $M_{Pl}$ for natural values of $\lambda_{\chi}$.  Therefore, in order for $G(\chi)$ to equal $M_{Pl}^{2}$ once $\chi = v_{\chi}$, $G(\chi)$ must increase very rapidly with $\chi$, with $G(v_{\chi}) \gg v_{\chi}^{2}$. It is common for the effective Planck mass due the non-minimal coupling in Higgs Inflation models, $M_{Pl\,eff}  = \sqrt{\xi_{\phi}} \phi$, to be much larger than the input field, by a factor of $\sim 10^{2}$ for metric Higgs Inflation and $\sim 10^{4}$ for Palatini Higgs Inflation. In the present case, the amplification is by a  larger factor, with $M_{Pl\,eff}/v_{\chi} \approx M_{Pl}/v_{\chi} \gae 10^{10}$. A second condition on $G(\chi)$ in this model is that it has to be negative at small $\chi$ but positive at larger $\chi$. Finally, in order to avoid a pole in the Einstein frame kinetic terms as $\chi$ increases from zero when $\phi = 0$ after inflation, we require that 
$M_{0}^{2} + G(\chi)$ is positive throughout. 
As a simple example, we can consider a non-minimal coupling $G(\chi)$ which behaves as 
\be{e69}   G(\chi) =  -\xi_{\chi} \chi^{2}   \;\;\;,\;\;\;  \chi < \chi_{c} \;\;\; ; \;\;\; G(\chi)  \approx  M_{Pl}^{2}   \;\;\;,\;\;\;  \chi > \chi_{c}   ~.\ee 
The Einstein frame potential for $\chi$ when $\phi = 0$ is then 
\be{e70} V_{E}(\chi) \approx \left(\frac{M_{Pl}}{M_{0}}\right)^{4} \frac{\lambda_{\chi}}{4} \left( \chi^{2} - v_{\chi}^{2} \right)^{2} \;\;\;,\;\;\;  \chi < \chi_{c}   \;\;\; ; \;\;\; V_{E}(\chi) = \frac{\lambda_{\chi}}{4} \left( \chi^{2} - v_{\chi}^{2} \right)^{2} \;\;\;,\;\;\;  \chi > \chi_{c}   ~,\ee
with a sharp decrease in the potential at $\chi = \chi_{c}$. This potential has a unique stable minimum at $\chi = v_{\chi}$.   
In order to avoid a pole in the Einstein frame kinetic terms we also require that $\xi_{\chi} \chi_{c}^{2} < M_{0}^{2}$.
In the absence of a more fundamental theory, the functional form of $G(\chi)$ is unknown. However, as long as the above conditions on $G(\chi)$ are satisfied, the specific form of $G(\chi)$ will not significantly affect the dynamics of inflation. 

An example of a relatively simple function of $\chi$ that can satisfy the behaviour in \eq{e69} is given by 
\be{e69a} G(\chi) = \frac{ M_{Pl}^{2} \left(1 - e^{-\chi^{2}/\chi_{c}^{2}} \right) \left(1 - A e^{-\chi^{2}/\chi_{c}^{2}} \right) }{ \left(1 + B e^{-\chi^{2}/\chi_{c}^{2}} \right)^{2} }   ~,\ee 
where $A > 1$ and $B \gg 1$. Then $G(\chi)$ satisfies
\be{e69b} G(\chi) \approx - \frac{M_{Pl}^{2}(A-1)}{B^{2}} \frac{\chi^{2}}{\chi_{c}^{2}}  \;\;;\;\; \chi^{2} \ll \chi_{c}^{2} ~,\ee
 \be{e69c} G(\chi) \approx  \frac{M_{Pl}^{2}}{B^{2}} e^{2 \chi^{2}/\chi_{c}^{2}}  \;\;;\;\; \chi_{*}^{2} \gg \chi^{2} \gg \chi_{c}^{2} ~\ee
and 
 \be{e69d} G(\chi) \approx M_{Pl}^{2}  \;\;;\;\; \chi^{2} \gg \chi_{*}^{2} ~,\ee 
where $\chi_{*} \approx \sqrt{\ln B}\, \chi_{c}$. Comparing with \eq{e69}, we have $M_{Pl}^{2}(
A-1)/B^{2} \approx \xi_{\chi} \chi_{c}^{2}$.  For example, $\chi_{c} = 10^{8} \GeV$, $\xi_{\chi}= 10^{4}$ and $A \sim 1$ gives $B \approx 10^{8}$ and $\chi_{*} \approx 4.3 \chi_{c}$. Therefore $G(\chi) \approx -\xi_{\chi} \chi^{2}$ until $\chi \approx \chi_{c}$, it then turns positive once $\chi > \chi_{c}$ and rapidly increases to $M_{Pl}^{2}$ at $\chi \approx \chi_{*} \approx 4.3 \chi_{c}$, becoming constant for $\chi > \chi_{*}$.  The requirement that $\xi_{\chi} \chi_{c}^{2} < M_{0}^{2}$ is satisfied if $M_{0} \gae 10^{10} \GeV$.

\section{Conclusions} 

 We have shown that it is possible for Higgs Inflation to occur at field values during inflation which are less than the instability scale of the unmodified Standard Model Higgs potential, if there is a phase transition at the end of inflation which increases the effective Planck mass in the Jordan frame.  This is equivalent to a suppression of the Higgs kinetic and potential term in the Einstein frame at the end of inflation. The classical predictions of the model are exactly the same as for conventional Higgs Inflation with the exception of the value of the Higgs field during inflation. We have presented a specific example of a model with a non-minimally coupled induced gravity sector which can account for Higgs Inflation with $\phi = 7.2 \times 10^{8} \GeV $ at $N = 55$, less than the Higgs instability scale $\Lambda \sim 10^{10} \GeV$. The induced gravity scalar $\chi$ is kept at a $\chi = 0$ minimum during inflation by a non-minimal coupling which is negative at small $\chi$. This minimum becomes unstable once Higgs Inflation ends and $\phi \rightarrow 0$, allowing the induced gravity phase transition to occur and the Jordan frame Planck mass to grow to its present value.  The model requires a  non-minimal coupling of the induced gravity scalar $\chi$ which is negative at small $\chi$ but becomes positive at large $\chi$. It must also increase very rapidly with increasing $\chi$ in order to reach a value equal to $M_{Pl}^{2}$ when $\chi$ reaches the minimum of its potential at $v_{\chi} \lae 10^{8} \GeV$.

Whilst the classical predictions of Higgs Inflation are unmodified, the quantum corrections to the Higgs potential can result in deviations from the classical predictions. Since the quantum corrections are determined purely by the Standard Model particle content, the corrections and their effect on the predictions of the model are well-defined and are independent of the specific model for the Planck mass transition. An analysis of quantum corrections, the resulting predictions for $n_{s}$ and $r$, and their dependence on the renormalisation frame, is discussed in a companion paper \cite{pt1c}.  

The idea of using the Higgs potential of the unmodified Standard Model to produce inflation is attractive. It would allow the only known scalar particle, the Higgs boson, to serve as the inflaton without requiring the ad hoc addition of sufficiently strongly coupled new particles to the Standard Model to stabilise the potential. The Planck mass phase transition mechanism proposed here would allow Higgs Inflation to be achieved using the unmodified potential. In the specific model for the phase transition that we have presented, it requires only the addition of a scalar that is not coupled to the Standard Model sector to produce the increase of the Planck mass. It is likely that the late-time Planck mass phase transition will also lead to as yet unknown phenomenological and cosmological possibilities.

\end{document}